# A 600 VOLT MULTI-STAGE, HIGH REPETITION RATE GAN FET SWITCH*

D. Frolov, H. Pfeffer, G. Saewert†, Fermilab, Batavia, USA

*Abstract*

Using recently available GaN FETs, a 600 Volt three-stage, multi-FET switch has been developed having 2 nanosecond rise time driving a 200 Ohm load with the potential of approaching 30 MHz average switching rates. Possible applications include driving particle beam choppers kicking bunch-by-bunch and beam deflectors where the rise time needs to be custom tailored. This paper reports on the engineering issues addressed, the design approach taken and some performance results of this switch.

## INTRODUCTION

Currently available gallium nitride (GaN) transistors are at least a factor of five times both lower in capacitance and faster in switching speed than MOSFETs. This technology enables drivers for particle beam deflecting devices to be built that achieve new performance levels in terms of speed. Although the development of this switch was targeted for driving a particular beam deflector at Fermilab, pulse generators in general employ switches with which high switching speed and high voltage are desired.

One requirement at Fermilab for which a hard-switching driver has been proposed is a fast beam chopper in the 2.1 MeV Medium Energy Beam Transport (MEBT) section of what is now referred to as PIP2-Injector Test facility at Fermilab [1]. This is to be a CW operating LINAC delivering unique beam patterns concurrently to numerous experiments. Bunch pattern creation will be done on a bunch-by-bunch basis by the MEBT chopper having 6.15 ns bunch spacing (162.5 MHz). Operational requirements include greater than 500 Volts on each traveling wave kicker plate, more than 80% beam chopped out, and particle removal to better than $10^{-4}$ of each chopped-out bunch [2]. The ramifications on a hard-switching circuit as the driver for this chopper are to have an absolute worst case 4.0 ns rise/fall time (5-95%), be able to switch at 81.25 MHz in 100 ns bursts at 1 μs intervals and handle average switching rates of more than 30 MHz when the PIP2 LINAC eventually operates CW.

Having the ability to build a switch in house is convenient for building deflection plate drivers that must ramp linearly at controlled rates rather than simply turn on as fast as possible. A version of this switch is under development for an electron beam profiler in the Fermilab Main Injector [3]. A pair of deflection plates needs to be ramped to plus and minus 500 V linearly in 18 ns to sweep an electron beam longitudinally across proton bunches.

A previous attempt had been made to build a fast switch for this PIP2 MEBT chopper using a different design approach. This paper reports on achieving better results by driving each FET stage individually and the employment of different GaN FETs rated for higher voltage.

\* Work supported by Fermi Research Alliance, LLC under Contract No. De-AC02-07CH11359 with the US Department of Energy.
† saewert@fnal.gov

## DESIGN REQUIREMENTS

This switch was designed to meet several requirements that include:
(1) Operation to voltages of at least 500 V,
(2) Turn-on transition times of 2 ns,
(3) The capability of switching at tens of megahertz CW and hopefully over 30 MHz,
(4) Pulse patterns of virtually any duty factor
(5) Pulse widths down to 2 ns at flat top.

Additionally, understanding that to operate at such high CW switching rates, the switch needs to be constructed of multiple FETs in series to share the power dissipation. So the design needed to include the ability to assure that multiple FETs can be made to be timed simultaneously and held to small fractions of a nanosecond. Development of this switch thus far were constructions of 2- and 3-FET assemblies.

Certainly lead inductance needs to be minimized. So this switch design needs to be physically as small as possible.

## DESIGN ISSUES

A number of design issues must be addressed besides switching fast. Foremost are switching losses. At tens of megahertz and 600 Volts these losses are high even for GaN to use only one FET as a switch. A FET's internal dissipation when discharging its own drain-to-source (Cds) capacitance is $0.5 C_{ds} V^2 f = 108$ W; where Cds = 20 pF, and f = 30 MHz switching CW. We are referring to parameters for the GS66502 GaN FET from GaN Systems, Inc., used in this design. Add to this the transition switching loss that arises during turn-on and turn-off. There are 36 Watts transition switching losses with a 4.0 ns rise time driving a 3.0 A load at 30 MHz CW. (These losses drop proportionately with rise time, however.) Not a switching loss, but each FET's drain-to-source $I^2R$ loss conducting 3 A DC is only 2 W.

It is interesting that total switching losses of a multi-FET in series are less than if the switch is comprised of only one FET. The reduction is Cds switching loss is evident from the expression for the total when three FETs are in series:

$$3 * \frac{1}{2} Cds \left(\frac{V}{3}\right)^2 f = \frac{1}{3} * \frac{1}{2} Cds V^2 f \ (Watts).$$

Where $V$ is the operating voltage, and $f$ is switching rate, the total is 1/3 compared when only using one FET. Share those losses among several GaN FETs, and cooling the FETs now becomes more manageable.

Common mode capacitance in the form of parasitic capacitance needs to be minimized between each FET stage and ground. Five Watts are dissipated in the FETs driving every picofarad of parasitic capacitance to ground at 30 MHz transitioning 600 V. That number comes down for the other stages, but each PCB is a couple of picofarads. Additionally, the FETs conduct this parasitic charging current on each voltage transition that only exacerbates the task of switching in 2 ns.

The circuit directly driving the GaN FET of this design is a discrete solution, since no commercial FET driver IC is suitable at these switching speeds. Power is required to drive FETS fast. The GS66502 input capacitance Ciss is low, however forced air cooling the Driver PCB is still required.

Timing shifts result if the circuitry's bias voltages vary under wide-ranging switching conditions, so only voltage regulators were used – no Zener diodes. A 500 kHz AC power supply system was designed to deliver power to each stage. Power consumption of each stage is about 2.5 W divided between the transformer, the several DC regulators and the GaN FET driver circuit.

Trigger signals delivered to each stage must have very high transient immunity. The worst case is 300 kV/µs for the stage transitioning 600 V in 2 ns. Opto-couplers were not used. Those having low enough jitter were deemed just too physically big for the closely space boards and the added parasitics to ground. Transformers were used to communicate triggers to each board and wound having well under 1 pF common mode.

## SWITCH CIRCUIT TOPOLOGY

Figure 1 shows the switch system block diagram. The figure shows a high-side switch configuration. However, the individual GaN FET and Driver stages are identical and are totally isolated from ground. Switches can thus be configured without modifications in the traditional topologies: as a high-side, low-side or bipolar switch. Parasitic capacitance to ground, lead inductance through the switch and trigger driving ability limit the number that can practically be put in series.

Triggering is by way of set-reset pulses issued by the Trigger Generator. This board divides the input pulse waveform into narrow set and reset pulses coincident with the leading and trailing edges. The volt-seconds on the trigger transformers is kept very small keeping them physically small and common mode capacitance very low while allowing for any pulse width to be generated. Output pulses to 600 V even only 2 ns wide are produced.

Latches on each Driver stage reconstruct the desired waveform. The critical timing is adjusted and fixed on each Driver board. Turn-on and -off delay and rise time are matched between the boards to about 100 ps.

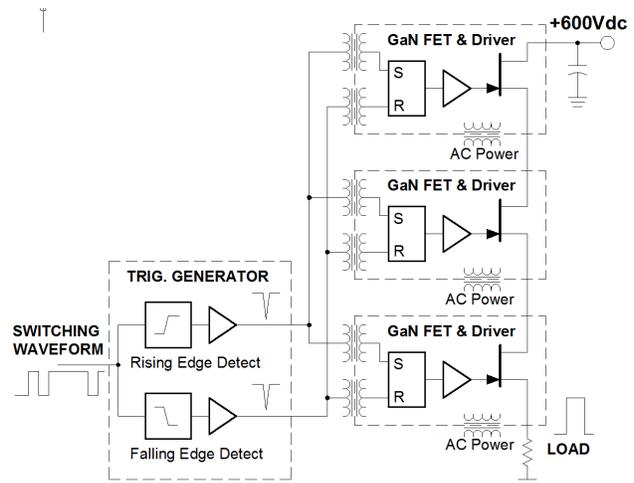

Fig. 1. Switch system diagram. GaN FET and Driver circuits are identical.

## SWITCH PERFORMANCE

Fig. 2 shows this 3-stage switch response, low-side configured, driving a 185 Ohm load to 550 V. The turn-on time is 2.0 ns (10-90%). The turn-off time is defined by the RC time constant formed by load resistance and switch output capacitance; it is longer than it takes for the switch to actually open. Turn-off time for single-switch topologies can be decreased by installing a speed-up network on the output.

One requirement for this switch in driving the PIP2 MEBT chopper is to switch at 81.25 MHz to kick out alternate bunches for 100 ns intervals. Figure 3 demonstrates this with a 3-stage switch driving the 200 Ohm chopper load. This switch, configured as a high-side switch, does include a speed-up network that decreases its turn-off time.

However, also evident are variations among the individual pulses arising from present limitations of the trigger transformers and are presently being resolved. The multi-FET switch per se is not the limiting factor.

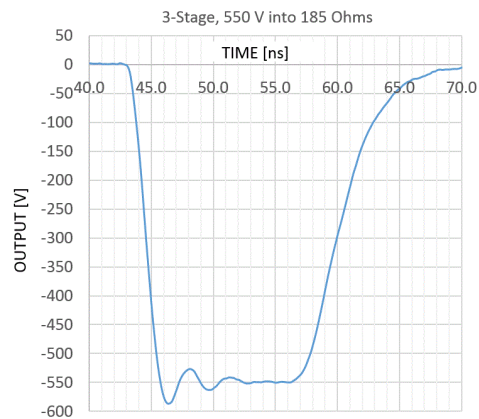

Fig. 2. 3-stage Low-side switch driving a 185 Ohm load to 550 V.

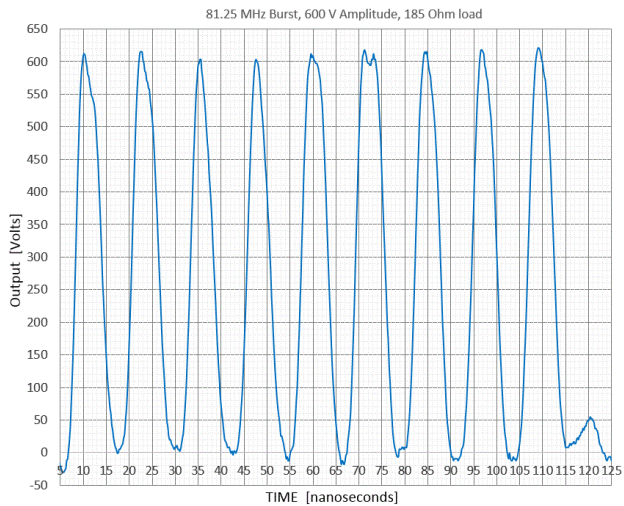

Fig. 3. 81.25 MHz burst to 600 V driving a 3 A load.

Efforts were made to determine whether this GaN FET and in this switch design are capable of operating at tens of megahertz. The junction temperature of the GaN FETs measured at 61 ºC when a 3-stage switch was operated steady-state at 2 MHz CW operating at 500 V into 185 Ohms. It was only air-cooled at about 1000 fpm. That was one semi-fast switching condition.

Other tests correlated DC power dissipation with steady-state switching at 200 V for various switching rates of a single FET. A high-side configuration was used to include all switching losses, and the same cooling was applied. Thermal images of the FET junction temperature correlated power dissipation when switching, and these figures matched power dissipation calculations that take all operating conditions into account including parasitics. Extrapolations predict 84 W dissipation distributed among the FETs in a 4-stage switch operating at 540 V at 35 MHz driving 185 Ohms. ANSYS analysis of one proposal showed there is at least one cooling scheme that is feasible.

What does happen beginning at FET junction temperatures above 55 ºC is turn-on and turn-off delays shift rather proportional with temperature. The direction of shift seems to correspond with GaN FET gate threshold voltage positive temperature coefficient. Given that this is predicable, it is conceivable that compensation can be included in the waveform generator knowing the GaN FET's thermal time constants, which are obtainable.

## PHOTO

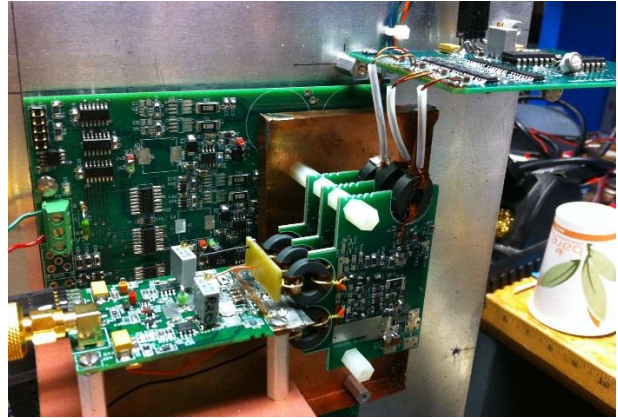

Fig. 4. Switch assembly. Trigger generator board is on the lower left, the 3-FET switch is center-right and AC power delivery PCB is upper-right. The load is not shown in this photo.

## SUMMARY

A switch has been developed and demonstrated to operate at 600 V having 2 ns rise and fall times driving a 3 A load. Switch assemblies consists of multiple GaN FET stages connected in series to reduce FET power dissipation when switching at megahertz rates at full voltage. Being a DC switch means it is capable of delivering a wide variety of pulse widths and duty factors without producing base line shifts or pulse droop. A switch delivers pulse widths from narrower than 2 ns at full voltage to indefinite. Switches were built and tested with 3 stages.

## ACKNOWLEDGEMENTS

The authors acknowledge the support of Fermilab's laboratory directed R&D program for its support of the majority of this development effort. Also appreciated is the assistance of Jeff Simmons for his craftsmanship.